\documentclass[twocolumn,showpacs,preprintnumbers,amsmath,amssymb]{revtex4}
\usepackage{graphicx}
\usepackage{dcolumn}
\usepackage{bm}
\usepackage{color}

\newcommand {\be}{\begin{equation}}
\newcommand {\ee}{\end{equation}}
\newcommand {\bea}{\begin{eqnarray}}

\newcommand {\eea}{\end{eqnarray}}

\begin{document}

\title{Deduction of an invariant-mass spectrum $M(\Sigma\pi)$ for $\Lambda(1405)$ with mixed  $T_{ \Sigma\pi \leftarrow K^-p}$ and $T_{ \Sigma\pi \leftarrow \Sigma\pi}$ from Hemingway's data on the   $\Sigma^+ (1660) \rightarrow \Lambda(1405) + \pi^+ \rightarrow (\Sigma\pi)_{I=0} + \pi^+$ processes}

\author{Maryam Hassanvand$^{1,2}$, Yoshinori Akaishi$^{1,3}$, Toshimitsu Yamazaki$^{1,4}$}

\affiliation{$^{1}$ RIKEN, Nishina Center, Wako, Saitama 351-0198, Japan}
\affiliation{$^{2}$ Department of  Physics, Isfahan University of Technology, Isfahan 84156-83111, Iran}
\affiliation{$^{3}$ College of Science and Technology, Nihon University, Funabashi, Chiba 274-8501, Japan}
\affiliation{$^{4}$ Department of Physics, University of Tokyo, Bunkyo-ku, Tokyo 113-0033, Japan}

\thanks{} 
\date{\today, expanded version of Phys. Rev. C, in press}

\begin{abstract}
We formulated the $\Lambda(1405)$ (abbreviated as $\Lambda^*$) $\rightarrow (\Sigma\pi)^0$ invariant-mass spectra produced in the $K^- + p \rightarrow \Sigma^+ (1660) + \pi^-$, followed by $\Sigma^+ (1660) \rightarrow \Lambda(1405) + \pi^+ \rightarrow \Sigma\pi + \pi^+$, processes at $p(K^-) = 4.2$ GeV/$c$,  in which both the incident channel for a quasi-bound $K^-p$ state and its decay process to $(\Sigma \pi)^0$ were taken into account realistically. We calculated $M(\Sigma \pi)$ spectral shapes using mixed transition matrices,  $T_{21} = T_{ \Sigma\pi \leftarrow K^-p}$ and $T_{22} = T_{ \Sigma\pi \leftarrow  \Sigma\pi}$, for various theoretical models involving $\Lambda^*$. The asymmetric spectra were compared to old experimental data of Hemingway, and it was found that the  mixing of the two channels,  written as $(1-f) \, T_{21} + f\, T_{22}$, gave a better result than considering the individual channels, yielding $f = 0.376_{-0.019}^{+0.021}$, $M(\Lambda^*) = 1406.6_{-3.3}^{+3.4}$ MeV/$c^2$ and $\Gamma = 70 \pm 2$ MeV, nearly consistent with the 2014 PDG values.
\end{abstract}


\maketitle

\section{Introduction}

      Historically, in 1959 Dalitz and Tuan \cite{Dalitz59} predicted the existence of a strange quasi-bound  state of $K^- + p \rightarrow \Sigma + \pi$ with $I=0$ in their analysis of experimental $\bar{K}N$ scattering data. In 1961 its experimental evidence was found from the  
mass spectrum, $M((\Sigma\pi)^0)$, in the reaction $K^- + p \rightarrow (\Sigma \pi)^0 + \pi^+ \pi^-$ at $p_{K^-} = 1.15$ GeV/$c$ \cite{L1405}. The resonant state of $\Lambda(1405)$ with $J = 1/2, I = 0, S=-1$, called $\Lambda(1405)$, is located below the $\bar{K}N$ threshold, and decays to $\Sigma \pi$.   After half a century, this state has been certified as a four-star state in Particle Data Group data \cite{PDG12}. According to Dalitz and Deloff \cite{Dalitz-Deloff}, from an M-matrix fit to experimental data of Hemingway \cite{Hemingway}, 
the mass and width of this resonance were obtained to be $M = 1406.5 \pm 4.0$ MeV/$c^2$ and $\Gamma = 50 \pm 2$ MeV. It is interpreted as a quasi-bound state of $\bar{K}N$ coupled with continuum state of $\Sigma \pi$. The 27-MeV binding energy of $K^- + p$ indicates a strongly attractive $\bar{K} N$ interaction, and a series of deep and dense $\bar{K}$ nuclear systems were predicted based on the $K^-p$-$\Sigma\pi$ coupled-channel calculations \cite{Akaishi:02,Yamazaki:02,Yamazaki:04,Dote:04a,Dote:04b,YA07PRC}. In the mean time, chiral dynamics theories suggested two poles  \cite{JidoNPA03,HW08} in the coupled $\bar{K}N - \Sigma\pi$ scheme, to which counter arguments were given \cite{Akaishi:10}. In the double-pole hypothesis the $\bar{K}N$ attraction mainly arises from the upper pole lying around 1420 MeV/$c^2$ or higher, and thus becomes much weaker, and may thus contribute only to shallow $\bar{K}$ bound states. The question as to whether the $K^- p$ bound state is deep or shallow is of great importance from the viewpoints of kaon condensation \cite{Kaplan-Nelson, Brown94}, but still remains controversial. Experiments of Braun {\it et al.} at CERN \cite{Braun77} and of Zychor {\it et al.} at COSY \cite{Zychor08} provided some interesting data, but they are statistically poor. 
More recently, Esmaili {\it et al.} \cite{Esmaili10,Esmaili11} analyzed old bubble-chamber data of stopped-$K^-$ on $^4$He \cite{Riley} with a resonant capture process, and found the best-fit value to be $M = 1405.5 ^{+1.4} _{-1.0}$. Hassanvand {\it et al.} \cite{Hassanvand13} analyzed recent data of HADES on $p + p \rightarrow p + K^+ + \Lambda(1405)$ \cite{HADES12b}, and subsequently deduced  $M = 1405^{+11}_{-9}$ MeV/$c^2$ and $\Gamma = 62 \pm 10$ MeV.  Now, the new PDG values \cite{PDG14} have been revised to be $M = 1405.1 ^{+1.3}_{-1.0}$ and $\Gamma = 50.5 \pm 2.0$, upon adopting the consequences of these  analyses. Concerning the most basic $\bar{K}$ bound state, $K^-pp$ predicted in \cite{Yamazaki:02,YA07PRC}, experimental evidence for deeply bound states has been obtained by FINUDA \cite{FINUDA}, DISTO \cite{DISTO} and J-PARC E27 \cite{E27}. 
 
In the present paper we provide a theoretical formulation to analyze the old experimental data of Hemingway at CERN in the reaction $K^- p \rightarrow \Sigma^+ (1660) + \pi^- \rightarrow \Lambda(1405) + \pi^+ + \pi^- \rightarrow (\Sigma\pi)_{I=0} + \pi^+ +\pi^-$ processes at 4.2 GeV/$c$. The intermediate resonance state $\Sigma^+ (1660)$ was well selected in the initial reaction channel of $K^- + p \rightarrow \Sigma^+ (1660) + \pi^-$. 
This data has been analyzed by many theoreticians, but ended with unsatisfactory consequences. One of the reasons might be because they did not examine the nature of the transitions from $\Sigma^+(1660)$ in terms of {\it both} $K^- + p \rightarrow \Sigma + \pi$ (expressed by $T_{21} = T_{ \Sigma\pi \leftarrow K^-p}$) 
{\it and}  $\Sigma + \pi \rightarrow \Sigma + \pi$ (expressed by $T_{22} = T_{ \Sigma\pi \leftarrow  \Sigma\pi}$). There were uncertainties in the selection between $T_{21}$ and $T_{22}$, and the data were often fitted by only $T_{22}$. Also, fitting was made for a Breit-Wigner shape, which is not justified because the resonance zone exceeds the kinematically allowed limits \cite{AMY08,Hassanvand13}. 

\begin{figure}
 \center\includegraphics[width=9cm]{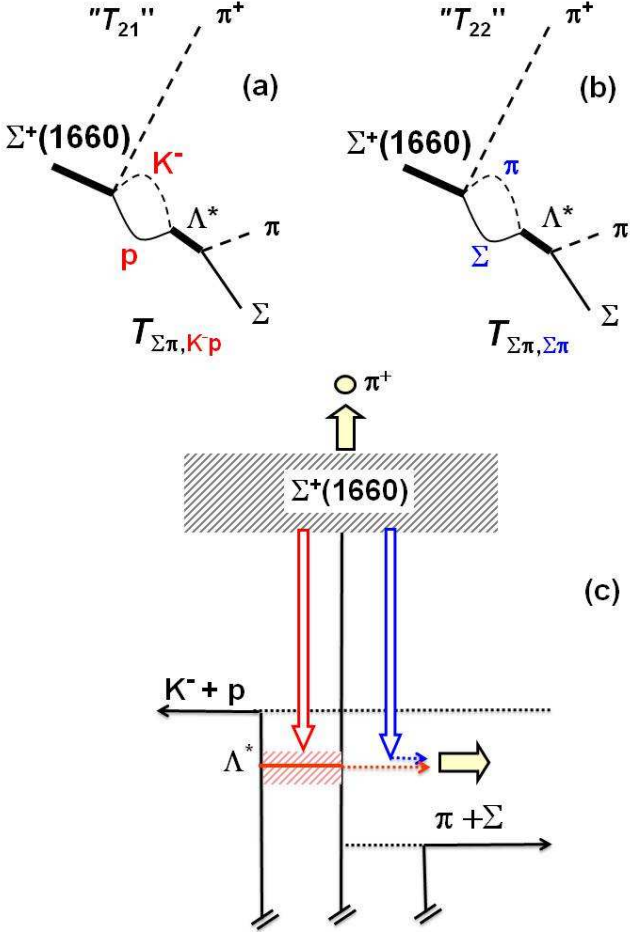}
    \caption{(Color online) Diagrams for the $\Sigma^+(1660)\rightarrow\pi^++\Lambda^*\rightarrow\pi^++(\Sigma \pi)^0$ reaction for (a) the process through $T_{\Sigma \pi\leftarrow K^-p}$ ($T_{21}$) and (b) the process through $T_{ \Sigma\pi \leftarrow  \Sigma\pi}$ ($T_{22}$) channels. (c) Schematic picture of the coupled-channel model in which the $K^-p$ quasi-bound state is treated as a Feshbach resonance, including two channels, $\bar{K}N$ and $ \Sigma\pi$.}\label{Fig1}
\end{figure}  

We show in Fig. \ref{Fig1} the level scheme of the decay of $\Sigma^+(1660)$ into a $K^-p$ quasi-bound state embedded in the continuum of $(\Sigma + \pi)_{I=0}$. There are two possible diagrams (a) and (b), which correspond to $T_{21}$ and $T_{22}$, respectively.  
Regarding the formation process, it is not obvious which of $T_{21}$ or $T_{22}$ is responsible for the Hemingway process that undergoes through $\Sigma^+ (1660)$. We thus set up arbitrarily mixed transition matrices, $T_{21}$ + $T_{22}$, for any kind of the $\bar{K}N$ interaction model so as to find the best fit with the experimental data without any prejudice. In addition, it is vitally important to take care of the broad distribution of the $\Lambda(1405)$ resonance, whose mass ranges between $M(\Sigma) + M(\pi) = 1330$ MeV/$c^2$ and $M(K^-) + M(p) = 1430$ MeV/$c^2$. Under these conditions the resonance shape can never be of a symmetric Breit-Wigner type, but should be very much skewed. Here, we follow our former treatments \cite{Hassanvand13}.  
   
\section{FORMULATION}

A coupled-channel treatment of $\Lambda(1405)$ employed in this paper is described in \cite{AMY08,Hassanvand13}. We use a set of separable potential with a Yukawa-type form factor,
\begin{equation}
\langle \vec k'_i |v_{ij} | \vec k_j \rangle = g(\vec k'_i) \, U_{ij} \,g(\vec k_j),
\end{equation} \label{eq:U-mat}
\begin{equation}
g(\vec k) = \frac {\Lambda^2} {\Lambda^2 + \vec k^2},~ \Lambda = \frac {m_B c} {h},
\end{equation} \label{eq:gk}
with $\Lambda$ being a range parameter, depending on the mass of exchanged boson ($m_B$), 
and 
\begin{equation}
U_{ij} = \frac {1} {\pi ^2} \frac{\hbar ^2}{2\surd \mu_i \mu_j} \frac{1}{\Lambda} s_{ij},
\end{equation}
where $i(j)$ stands for the $\bar{K}N$ channel, 1, or the $\Sigma \pi$ channel, 2, and $\mu_i$ ($\mu_j$) is the reduced mass of channel $i (j)$, and  ${s_{ij}}'s$ are non-dimensional strength parameters. We obtain $s_{11}$ and $s_{12}$ from the $M$ and $\Gamma$ values of an arbitrarily chosen $K^- p$ state to be used to calculate the $\Sigma \pi$ invariant masses. It means that, in our model, the strength parameters depend on the binding energy and the width of $\Lambda(1405)$ state as explained in detail in ref.~\cite{AMY08}. In our coupled-channel model presented here, it is obvious that the properly determined two parameters, $s_{11}$ and $s_{12}$, for any value of $s_{22}$, can represent the resonance pole without loss of generality. Here, we adopt $s_{22}=-0.7$, which gives $U_{22}/ U_{11} =4/3$ for $\Lambda(1405)$ as in a "chiral" model, and $\Lambda = 3.9$ fm$^{-1}$. 
 
As described in \cite{Esmaili10} in detail, we treat the $K^- p $ quasi-bound state as a Feshbach resonance, and the coupled-channel transition matrix, 
\begin{equation}
\langle \vec k'_i |t_{ij} | \vec k_j \rangle = g(\vec k'_i) \, T_{ij} \,g(\vec k_j),
\end{equation} \label{eq:t-mat}
satisfies the following matrix equation
\begin{equation}
T_{ij} = U_{ij}+ \sum \limits_{i=l} U_{il} G_{l} T_{lj},
\end{equation}
 with a loop function $G_l$:
\begin{equation}
G_{l} = \frac{2\mu_l}{\hslash^2} \int d\vec{q}~ g(\vec{q}) \frac{1}{k_l^2 - q^2 +i\epsilon_0} ~g(\vec{q}).
\end{equation} 
The solution is given in a matrix form by
\begin{equation}
T = [1-UG]^{-1} U,
\end{equation} 
with
\begin{equation}
(UG)_{lj} =-s_{lj} \surd \frac{\mu_j}{\mu_l} \frac{\Lambda^2}{(\Lambda - i k_j)^2},
\end{equation}  
and $k_j$ is a relative momentum in channel $j$.

The transition matrix elements in this framework are $T_{11}$, $T_{12}$, $T_{21}$ and $T_{22}$, which constitute 
the  experimentally observable quantities below the $\bar{K}+N$ threshold, $(-1/\pi) {\rm Im}(T_{11})$, $|T_{21}| ^2 k_{2}$ and $|T_{22}| ^2 k_{2}$ where $k_{2}$ is the $\Sigma \pi$ relative momentum. The first term corresponds to the $\bar{K}N$ missing-mass spectrum and is proportional to the imaginary part of the scattering amplitude given in Fig.15 of Hyodo-Weise \cite{HW08}. The second term with $g^2(k_2)g^2(k_1)$ is a $\Sigma \pi$ invariant mass from the conversion process, $\bar{K}N \rightarrow \Sigma \pi$ (called in this paper as "$T_{21}$ invariant mass") which coincides with the $\bar{K}N $ missing-mass spectrum in the mass region below the $\bar{K}+N $ threshold through the following formula, as has been derived from an optical relation \cite{Esmaili10}, 
\begin{equation}
{\rm Im} T_{11} = |T_{21}|^2\, {\rm Im}(G_2).
\end{equation}  \label{eq:T11T21}

Therefore the observation of a $T_{21}$ invariant-mass spectrum is just the observation of the imaginary part of the scattering amplitude given in \cite{HW08}.
The third term with $g^4(k_2)$ is a $\Sigma \pi$ invariant-mass spectrum from the scattering process, $\Sigma \pi \rightarrow \Sigma \pi$ (called in this paper as "$T_{22}$ invariant mass"). Two observables of $\bar{K}N-\Sigma \pi$ coupled channels (as mentioned above $T_{11}$ channel is associated with $T_{21}$ by eq.~\ref{eq:T11T21}.9) calculated by Hyodo and Weise's chiral two-channel model \cite{HW08} and also in the framework of $\Lambda(1405)$ ansatz of Akaishi and Yamazaki \cite{Akaishi:02} represented in fig.1 (upper) of Esmaili's paper \cite{Esmaili11}. This figure shows that the two curves of the chiral model have peaks at different positions (1420 and 1405 MeV/$c^2$) but Akaishi and Yamazaki's $T_{21}$ and $T_{22}$ invariant mass spectra have peaks near 1405MeV/$c^2$. Within this theory the peak positions can be varied. The main purpose of this paper is to determine the peak position of these two channels by comparing with experimental observables.

The level scheme for $\Sigma^+(1660)\rightarrow \Lambda^*+\pi^+\rightarrow (\Sigma \pi)^0+\pi^+$ is shown in fig.~\ref{Fig1}. It proceeds to either the $K^- + p$ channel forming the $\Lambda^*$ resonance, then decaying to $\Sigma \pi$, which corresponds to the diagram (a) ($T_{21}$). Another process is to emit $\Sigma + \pi$, which form $\Lambda^*$, then decaying to $\Sigma + \pi$, as represented by $T_{22}$ in (b). Therefor there are two "incident channel"s to bring $\Lambda(1405)$ state: one is $K^- + p$ and the other is $\Sigma + \pi$. This picture was also shown by Geng and Oset \cite{Geng-Oset} in a different framework. In the mechanism given in the present paper, the resonance state $\Lambda^*$ is a Feshbach state, in which a quasi-bound $K^-p$ state is embedded in the continuum of $\Sigma \pi$. 
  
The theoretical framework for calculating the decay rate of $\Lambda(1405)$ to $(\Sigma \pi)^0$ was given in detail in \cite{AMY08, Akaishi:10}. To calculate the decay rate function, we take into account the emitted $\Sigma$ and $\pi$ particles realistically, following the generalized optical formalism in Feshbach theory \cite{Feshbach58}. The decay function, $G(x)$ with $x=M_{\Sigma \pi}$ being the invariant-mass, is not simply a Lorentzian but is skewed because the kinetic freedom of the decay particles is limited. The general form of $G(x)$ is given as
\begin{equation}
G(x) =  \frac{ 2(2 \pi)^5}{(\hslash c)^2} \frac{E_{\Sigma} E_{\pi}}{E_{\Sigma}+E_{\pi}} Re[k] \,| \langle k \, | \,t \,| k_0 \rangle  |^2.
\end{equation} \label{G(x)}
where $k_0$ and $k$ are the relative momenta in the initial and final states written as
\begin{equation}
\vec{k}_0 =\frac{c \surd \lambda (x,m_K,M_p)}{2\hbar x},
\end{equation} 
and 
\begin{equation}
\vec{k} =\frac{c \surd \lambda (x,m_{\pi},M_{\Sigma})}{2\hbar x},
\end{equation} 
with
\begin{eqnarray}
&&\lambda (x,m_1,m_2) = \\ \nonumber
&& (x + m_1 + m_2)(x + m_1 - m_2)\\ \nonumber
&&\times (x - m_1 + m_2)(x - m_1 - m_2).
\end{eqnarray}

 The kinematical variables in the c.m. framwork of the decay process of $T_{21}$ and $T_{22}$ channels are given in fig.~2 of ref.~\cite{Hassanvand13}.
 
  In this way the decay rate of $\Lambda(1405)\rightarrow \Sigma + \pi$ process via two so-called "incident channels" $K^-+p\rightarrow \Lambda(1405)$ and $ \Sigma+ \pi \rightarrow \Lambda(1405)$ as shown in fig.~\ref{Fig1}(a) and (b) is obtained. Equation~\ref{G(x)}.10 with eq.~\ref{eq:U-mat}.1 and \ref{eq:gk}.2 is completed and the invariant mass spectra of $T_{21}$ and $T_{22}$ channels calculated using eq.~\ref{eq:t-mat}.4. As we show in fig.~3 (c) in our previous work \cite{Hassanvand13}, these spectra do not depend on the incident energy of the Hemingway experiment,  $E(K^-)$=4.2 GeV, in the laboratory and for all values of  $E(K^-)$s the shape of the spectrum does not change. $G(x)$ is a unique function of $x=M_{\Sigma \pi}$ (invariant-mass) associated with $m_i$ (mass of $K^-$, $p$, $\Sigma$ and $\pi$ particles) and is bounded by the lower end ($M_l=M_{\Sigma}+M_{\pi}$=1328 MeV$/c^2$) and the upper end ($M_u=M_{p}+M_{K^-}$=1432 MeV$/c^2$).
 
It should be noted here that by changing the width of the spectrum, the position of the peak in $G(x)$ moves, and is different from the position of the pole ($M$=1405 MeV/$c^2$). In the next section we present the $G(x)$ function as $S(x; M, \Gamma)$ to obtain the $\chi^2$ value. $S(x; M, \Gamma)$ spectrum has a peak and a width in the region of the experimental data (1330-1430 MeV/$c^2$). We first consider the binding energy and the width of the pole as two free parameters and calculate the spectrum. Then we compare these theoretical curves to the experimental data using a $\chi^2$ test which gives us the degree of fitting as to how well our model actually reflects the data. In Section III we discuss some results of current model and compare them to Hemingway's experimental data.

\begin{figure*}
\includegraphics[width=8cm]{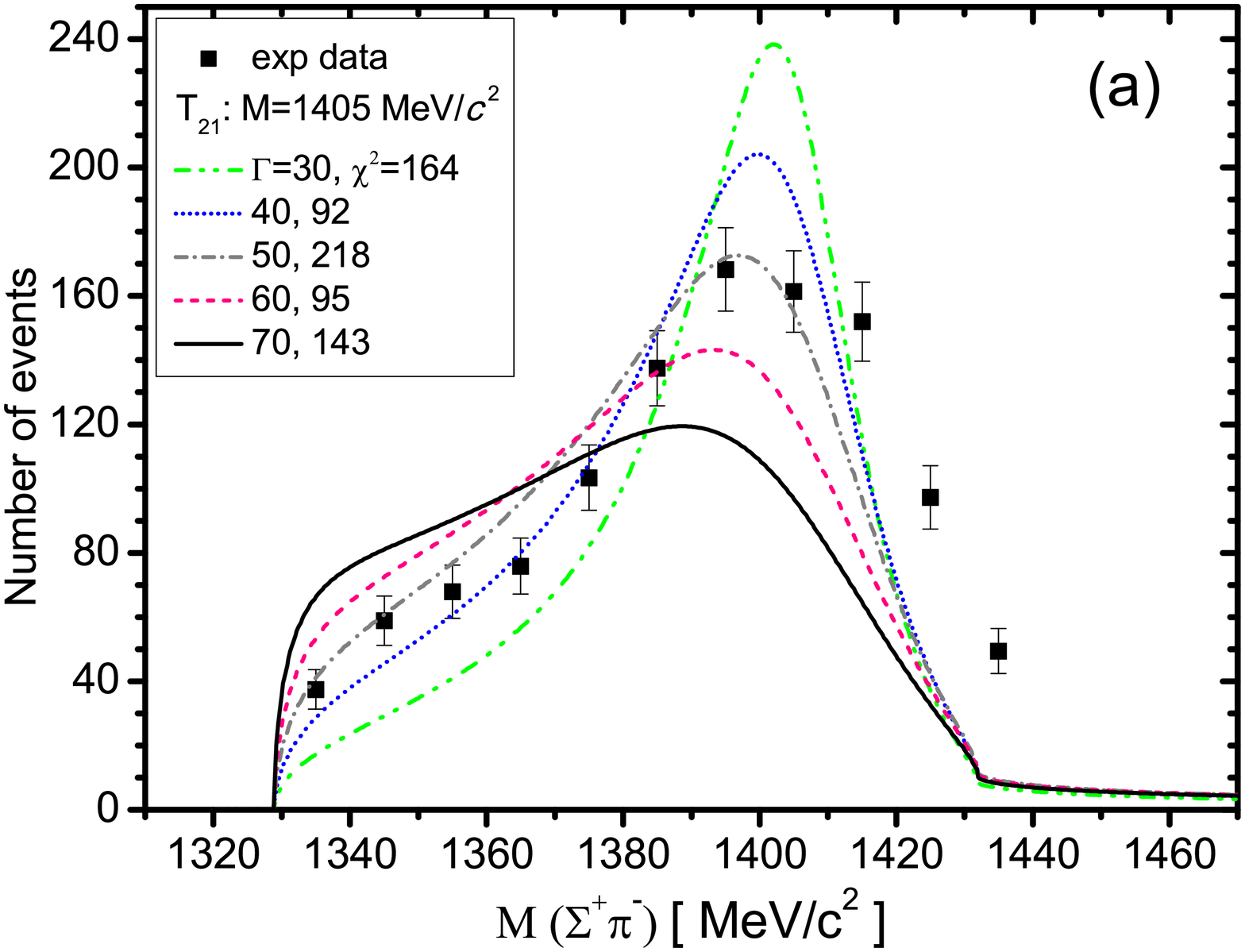}
\includegraphics[width=8cm]{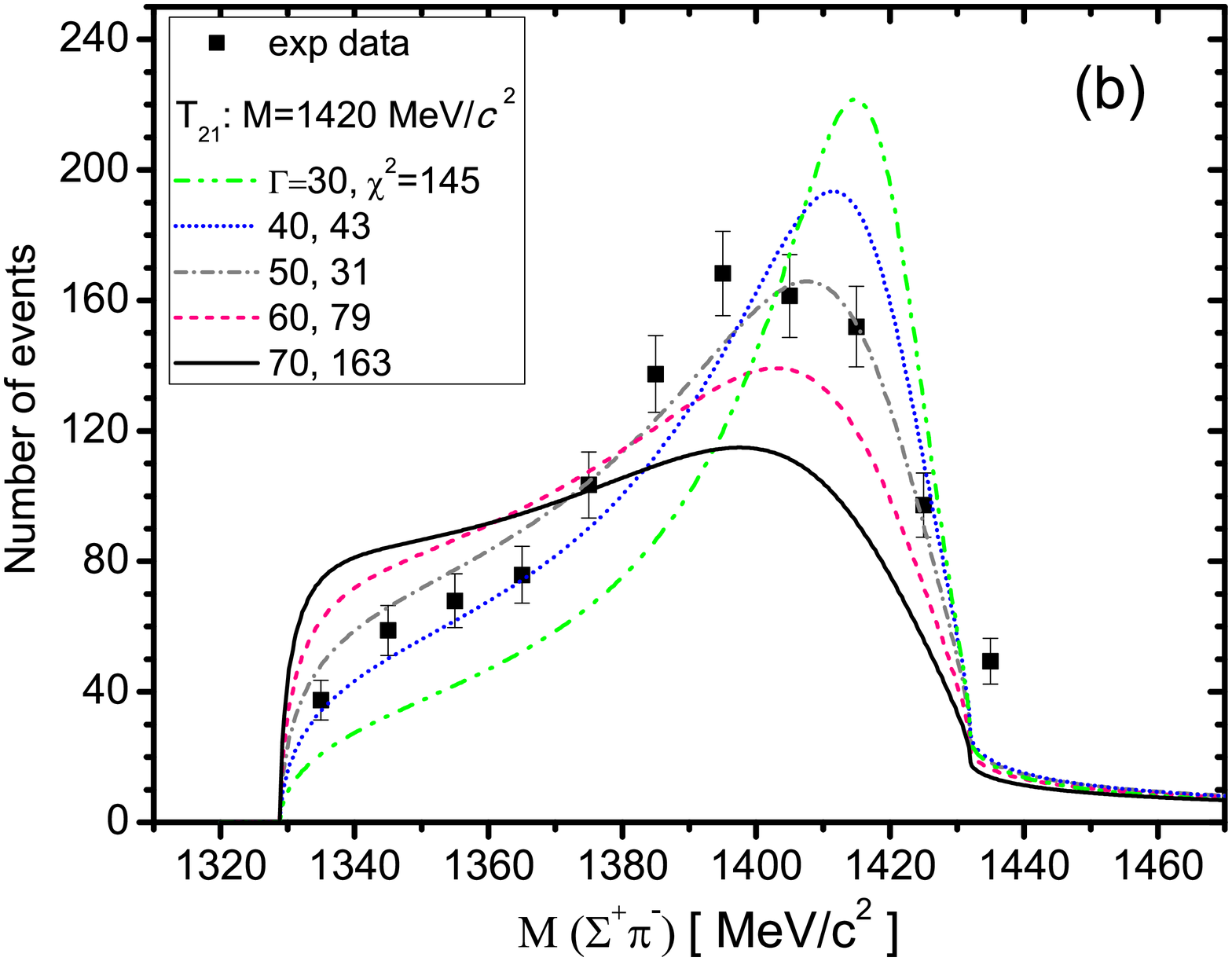}\\
\includegraphics[width=8cm]{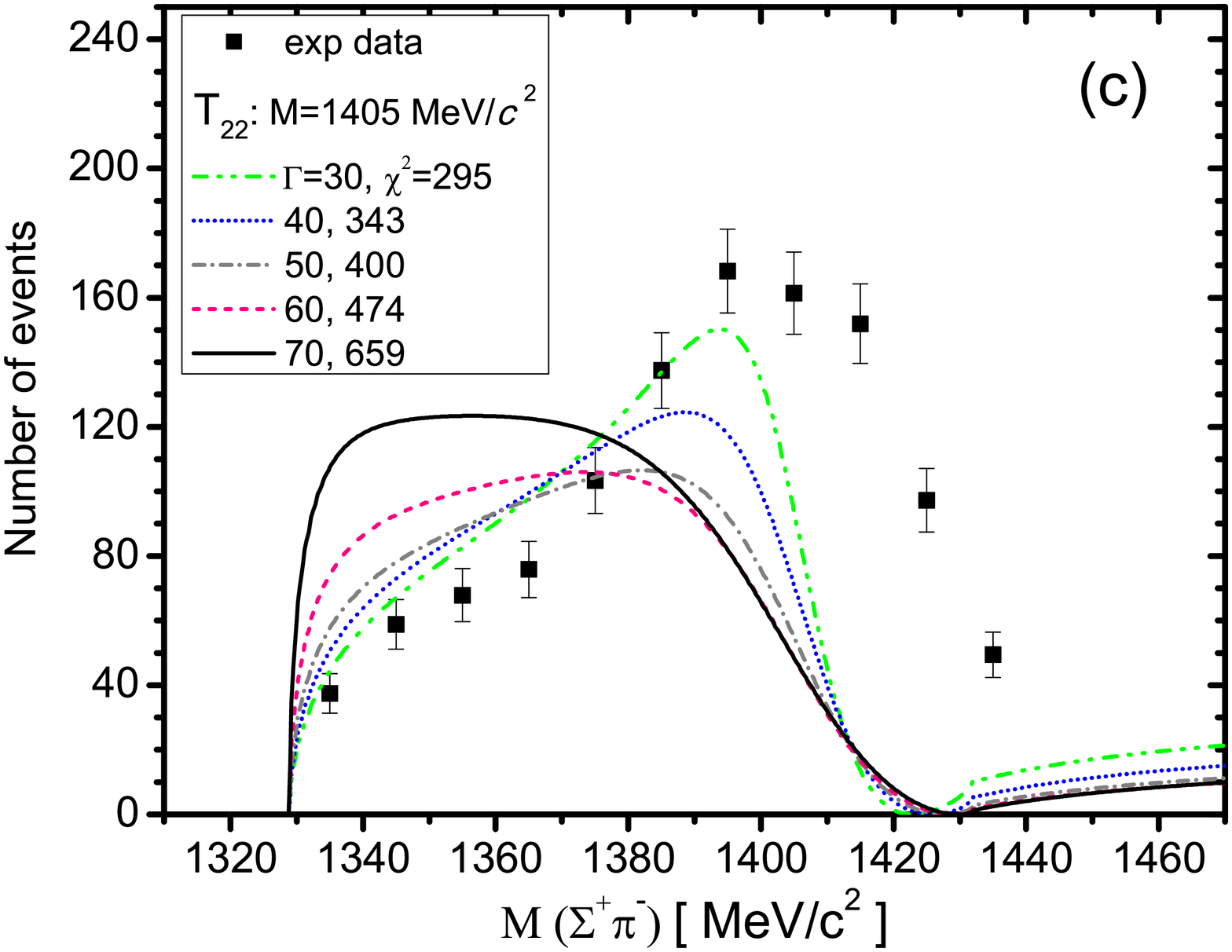}
\includegraphics[width=8cm]{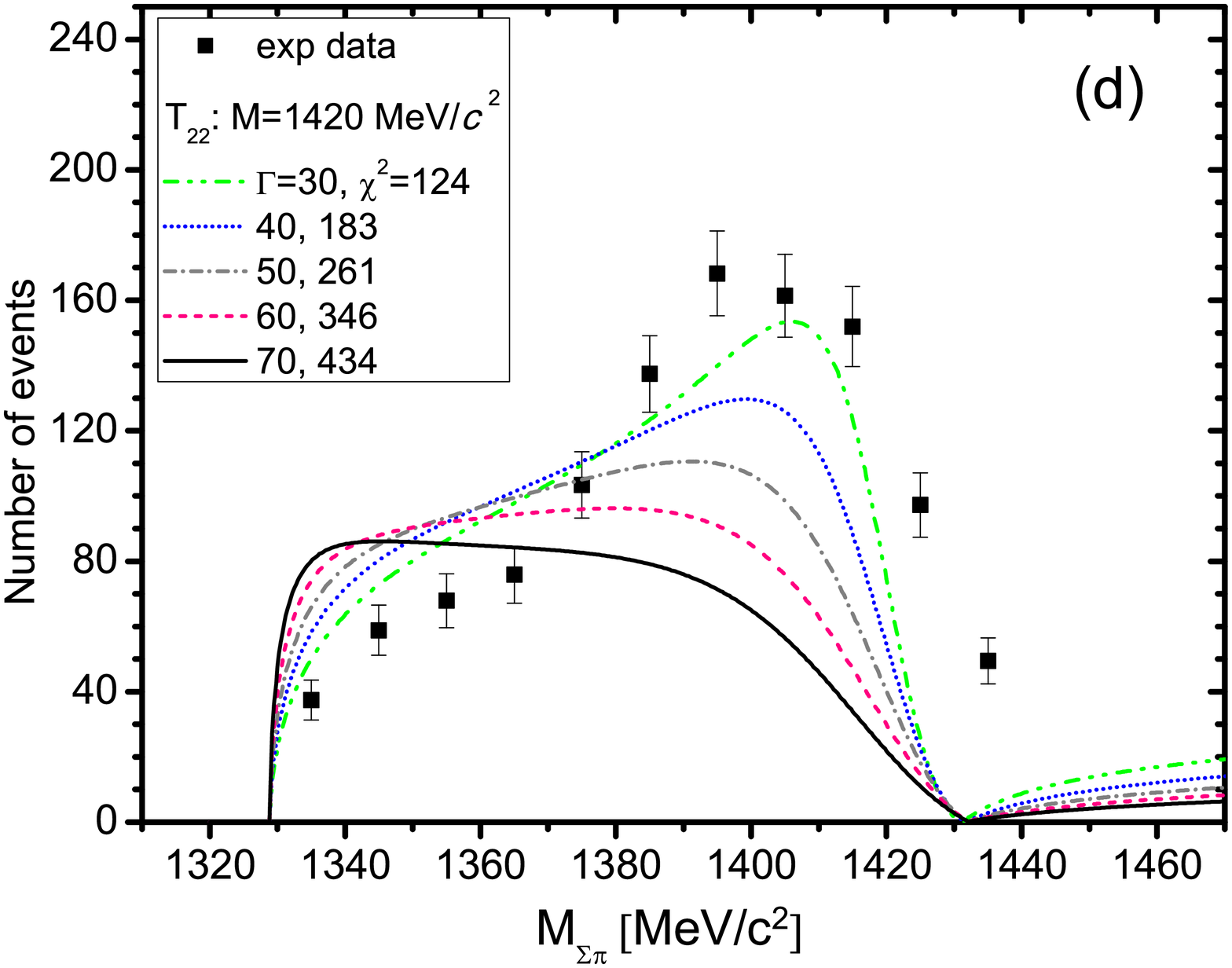}\\
\caption{(Color online) Spectra of the $|T_{21}|$ channel  (upper figures) and the $|T_{22}|$ channel  (lower figures) for a fixed $M_{\Lambda ^*}$ of 1405 (left frames) and 1420 (right frames) MeV/$c^2$ and for $\Gamma$ values from $30$ to $70$ MeV by $10$ MeV steps. The experimental values of Hemingway \cite{Hemingway} are shown by closed circles with error bars. The $\chi^2$ values from fitting to each curve are shown.  }\label{fig:T21_Mfix_1405}
\end{figure*}

\section{FITTING RESULTS}

The 11 points of Hemingway's data, ($N_i \pm \sigma_i$) for $i=1,...,n$, cover a mass spread from $M$=1330 MeV/$c^2$ ($\Sigma \pi$ threshold) to $M$=1430 MeV/$c^2$ ($K^-p$ threshold) \cite{Hemingway}. Previously, these data were fitted by a Breit-Wigner function, $K$-matrix calculation, and another model, namely, an extended cloudy bag model, as given in reference \cite{Hemingway}.
The Breit-Wigner function makes a poor fit to the data, yielding a mass and width of $\Lambda(1405)$ as $1391 \pm 1$ MeV/$c^2$ and $32 \pm 1$ MeV, while the $K$-matrix method results in $1411.4 \pm 2.0$ MeV/$c^2$ and $79.6$ MeV for its mass and width \cite{Dalitz-Deloff}.

The main purpose of the present paper is to fit the experimental data to our theoretical curves given in the preceding section so that the best fit with the least $\chi^2$ value can be deduced. The theoretical spectral curve, $S(x; M, \Gamma)$, is a function of the invariant-mass ($x=M_{\Sigma \pi}$) with the mass ($M$) and the width ($\Gamma $) of the $\Lambda^*$ resonance as parameters. Then, $\chi^2$ will be defined as 

\begin{equation}
\chi ^2 (M,\Gamma) = \sum (\frac{N_i - S(x_i; M, \Gamma)}{\sigma_i})^2,
\end{equation} 
where $N_i$ is the experimental data, $\sigma_i$ is the errors of the data.  

Using the $\chi^2$ method, the best possible fit has been obtained between the spectrum shape of the $\Sigma \pi$ invariant-mass, given from the $T_{21}$ ($K^-p \rightarrow \Sigma \pi $) and $T_{22}$ ($\Sigma \pi \rightarrow \Sigma \pi $) channels and Hemingway's experimental data.
For this work we faced a two-dimensional plane consisting of the mass of $\Lambda^*$ ($M$) and its width ($\Gamma$), so that by varying each of these parameters $\chi^2$ values could be obtained. Our purpose in this section is to describe how we obtained a pair of ($M_{\Lambda ^*}$,$\Gamma$) that give the minimum $\chi^2$.

We first overview how the theoretical $S(M, \Gamma)$ curves behave in comparison with the experimental data in Fig.\ref{fig:T21_Mfix_1405}, where calculated curves are shown together with the experimental data. The upper (or lower) two frames, specified with (a) (or (c)) and (b) (or (d)) in the figure, give curves for the $T_{21}$(or $T_{22}$) channel with assumed masses of $M = 1405$ MeV/$c^2$ (left) and 1420 MeV/$c^2$ (right), both with 5 different curves corresponding to assumed values of $\Gamma = $30, 40, 50, 60, and 70 MeV. The experimental data reveal a broad bump at around 1400 - 1420 MeV/$c^2$ with a long lower tail. A very characteristic feature of the theoretical curves is their asymmetric and skewed shapes, which can be understood in terms of the broad resonance located in the limited mass range. When $\Gamma$ is small, the curve shows a distinct peak at around the assumed mass, $M$, but the lower tail part cannot be accounted for. On the other hand, when $\Gamma$ becomes large, the lower tail component increases too much. At around $\Gamma \sim 50$ MeV a very crude agreement is attained, but the $\chi^2$ value is still large at around 50 compared with the expected $\chi^2$ value of $n_{\rm DF} = n - 3 \approx 8$. 

\begin{figure*} 
\includegraphics[width=8cm]{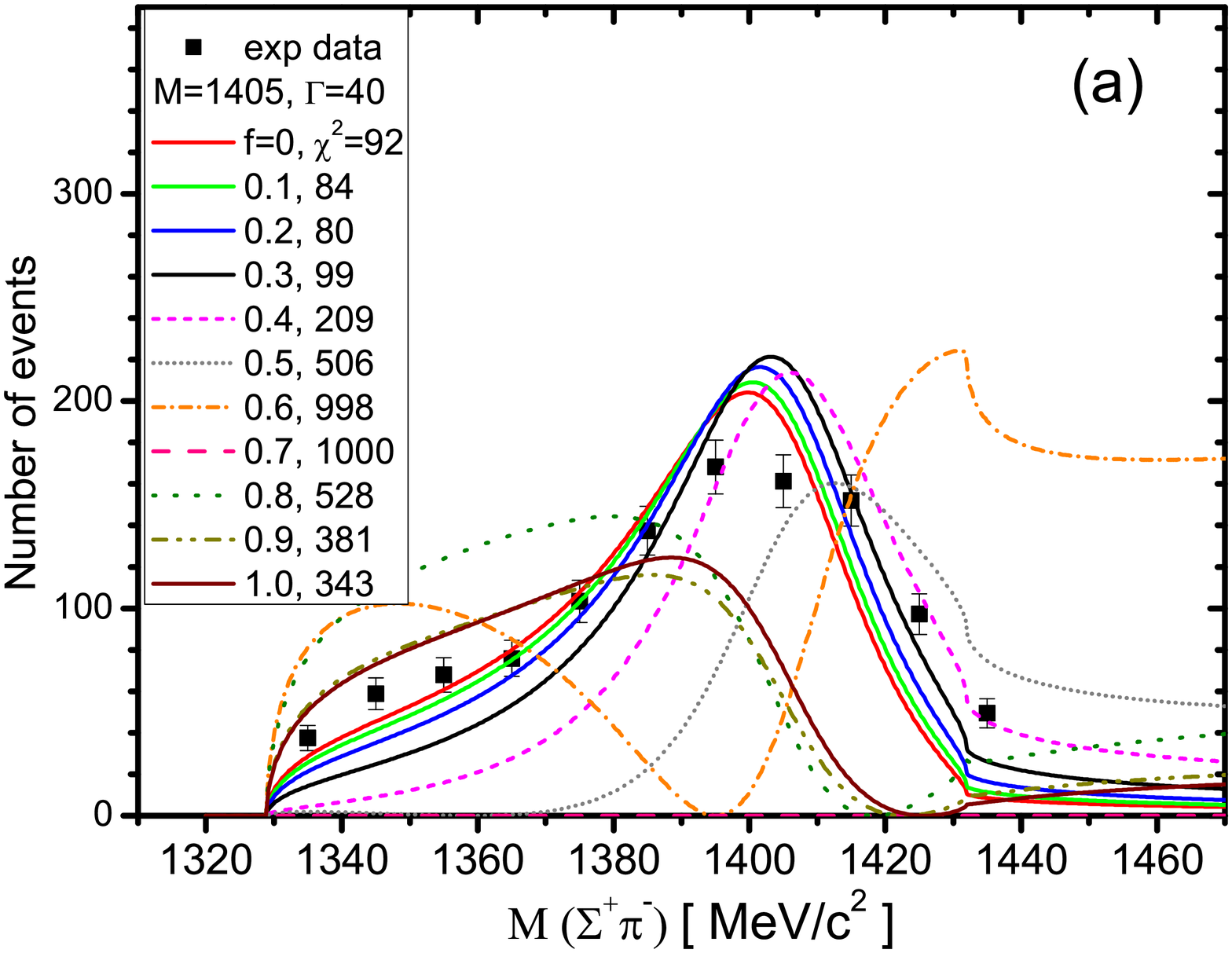}
\includegraphics[width=8cm]{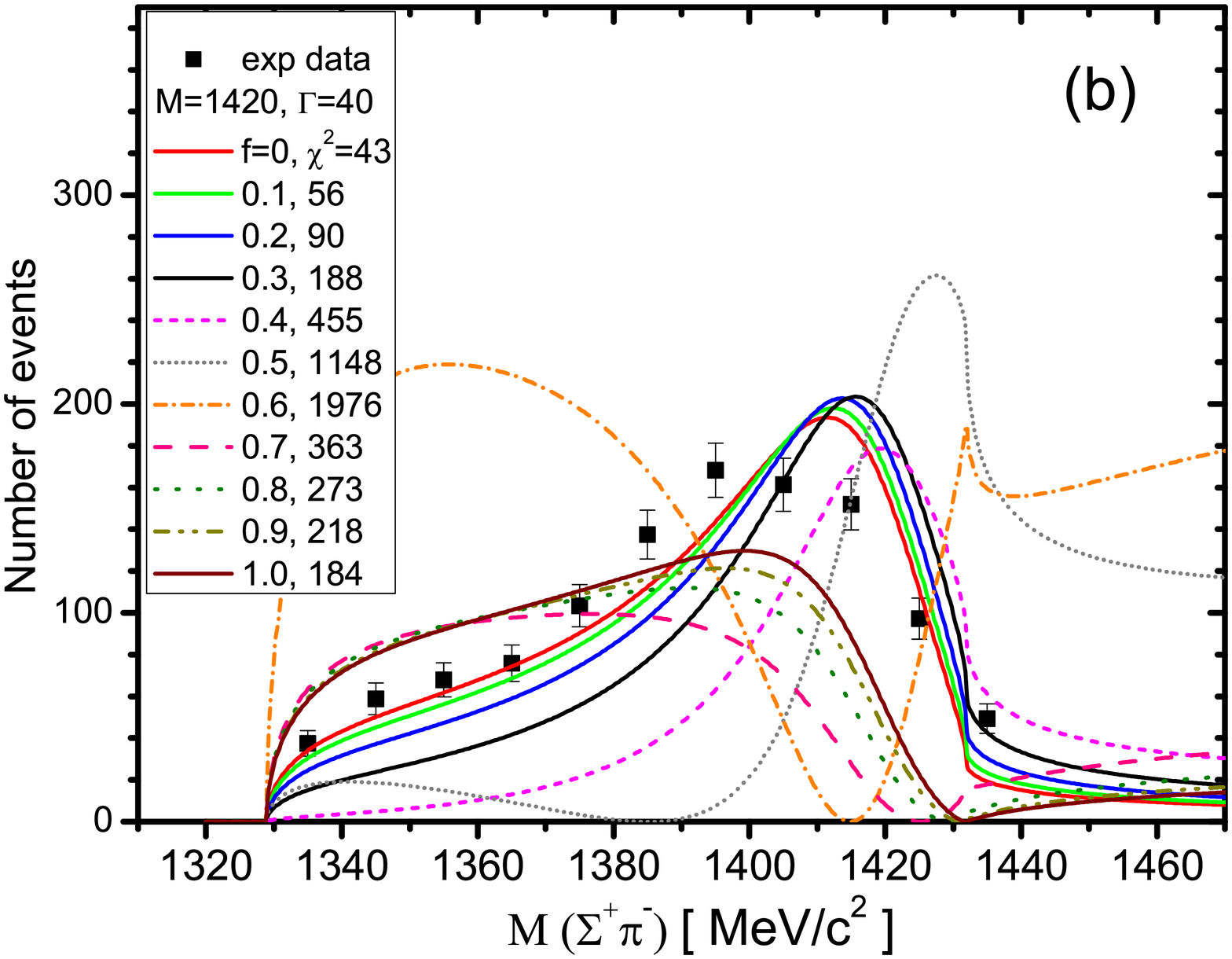}\\
\includegraphics[width=8cm]{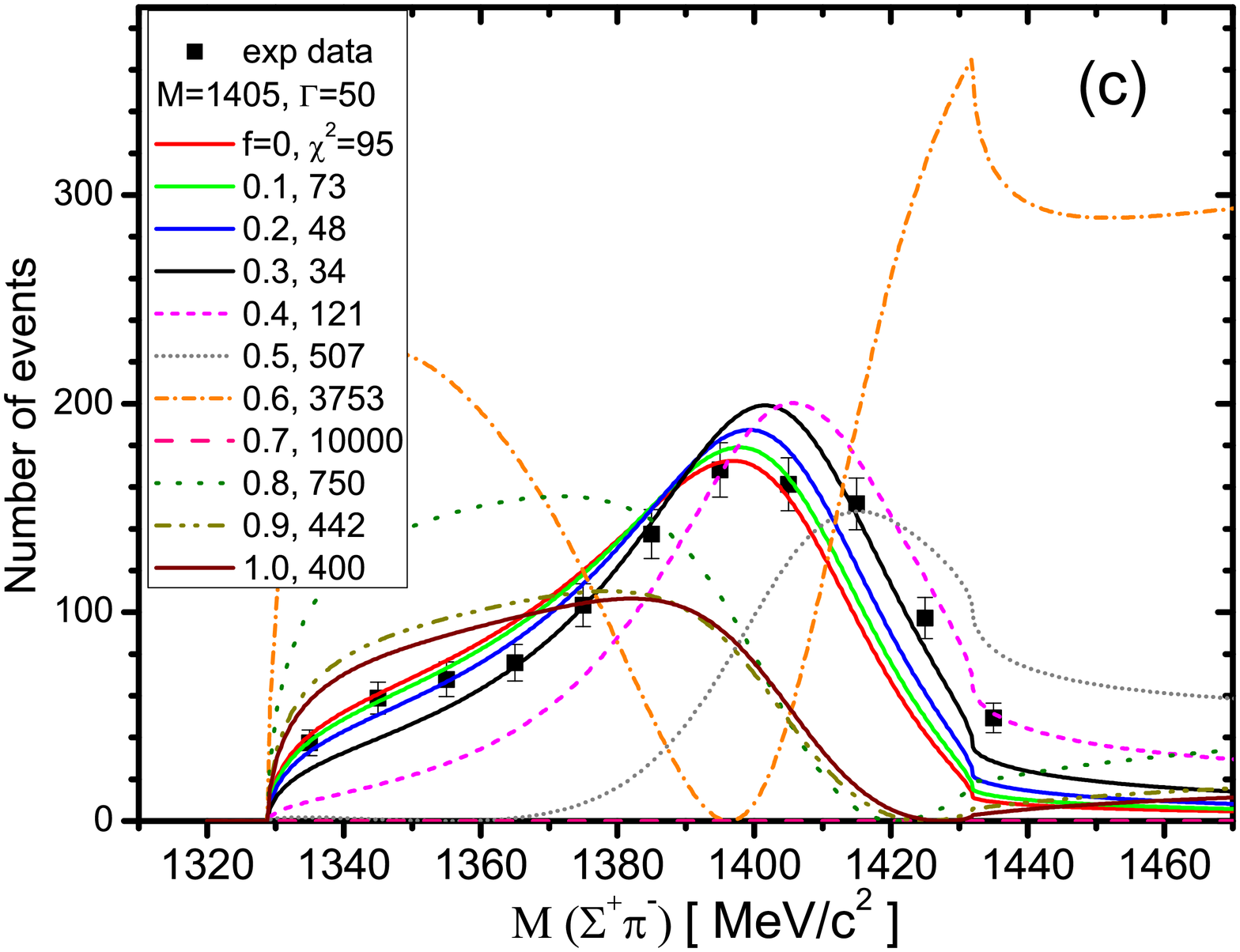}
\includegraphics[width=8cm]{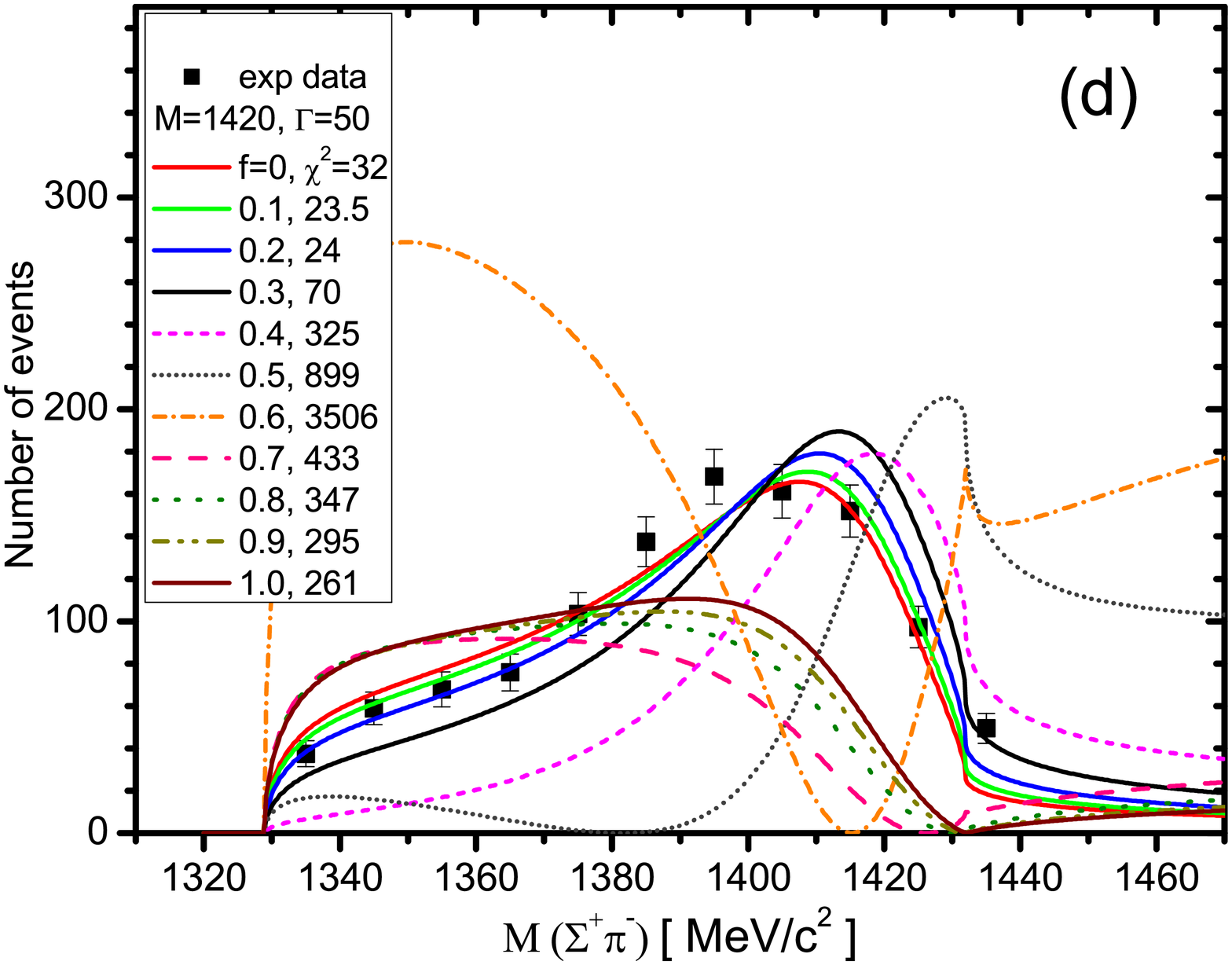}\\
\includegraphics[width=8cm]{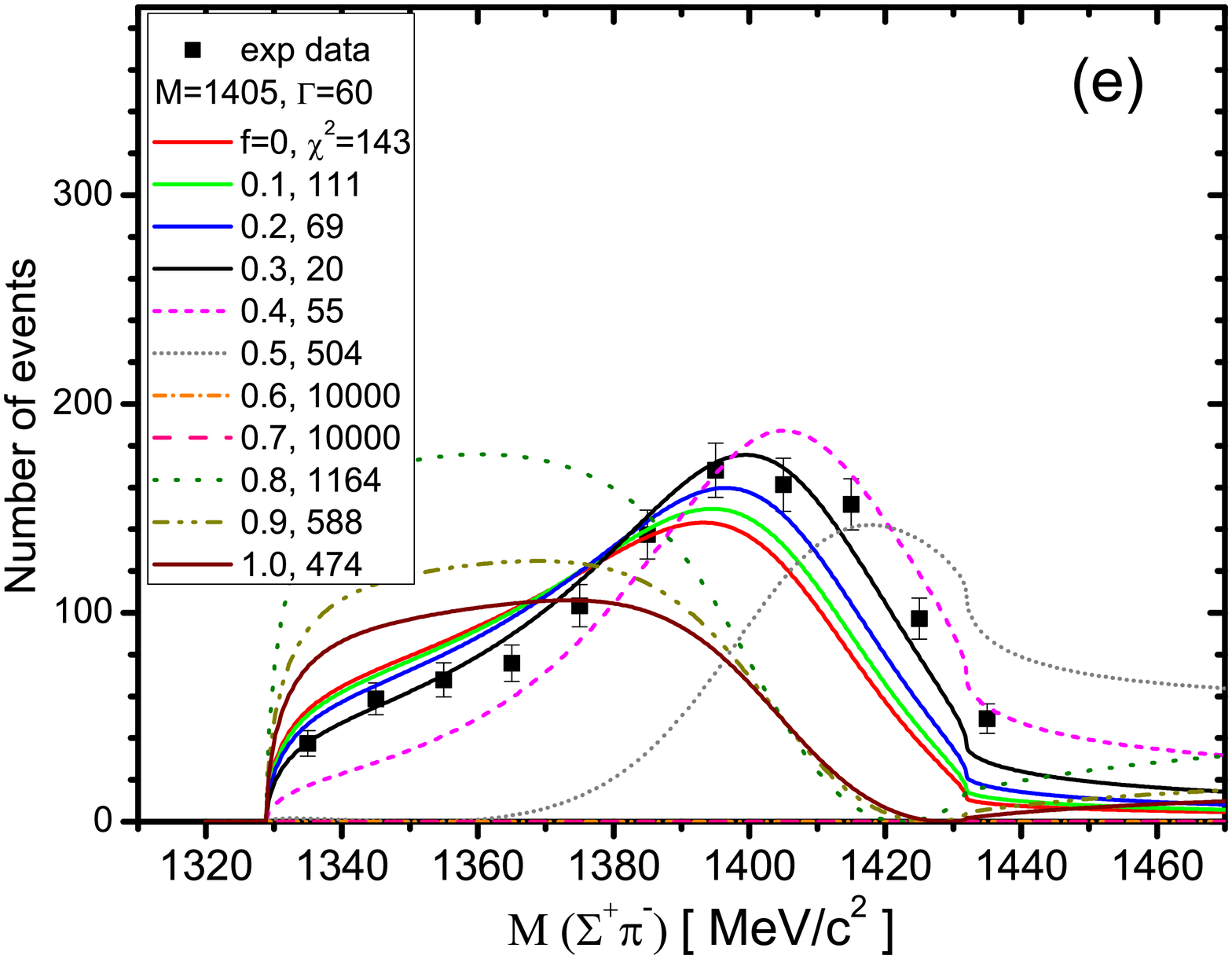}
\includegraphics[width=8cm]{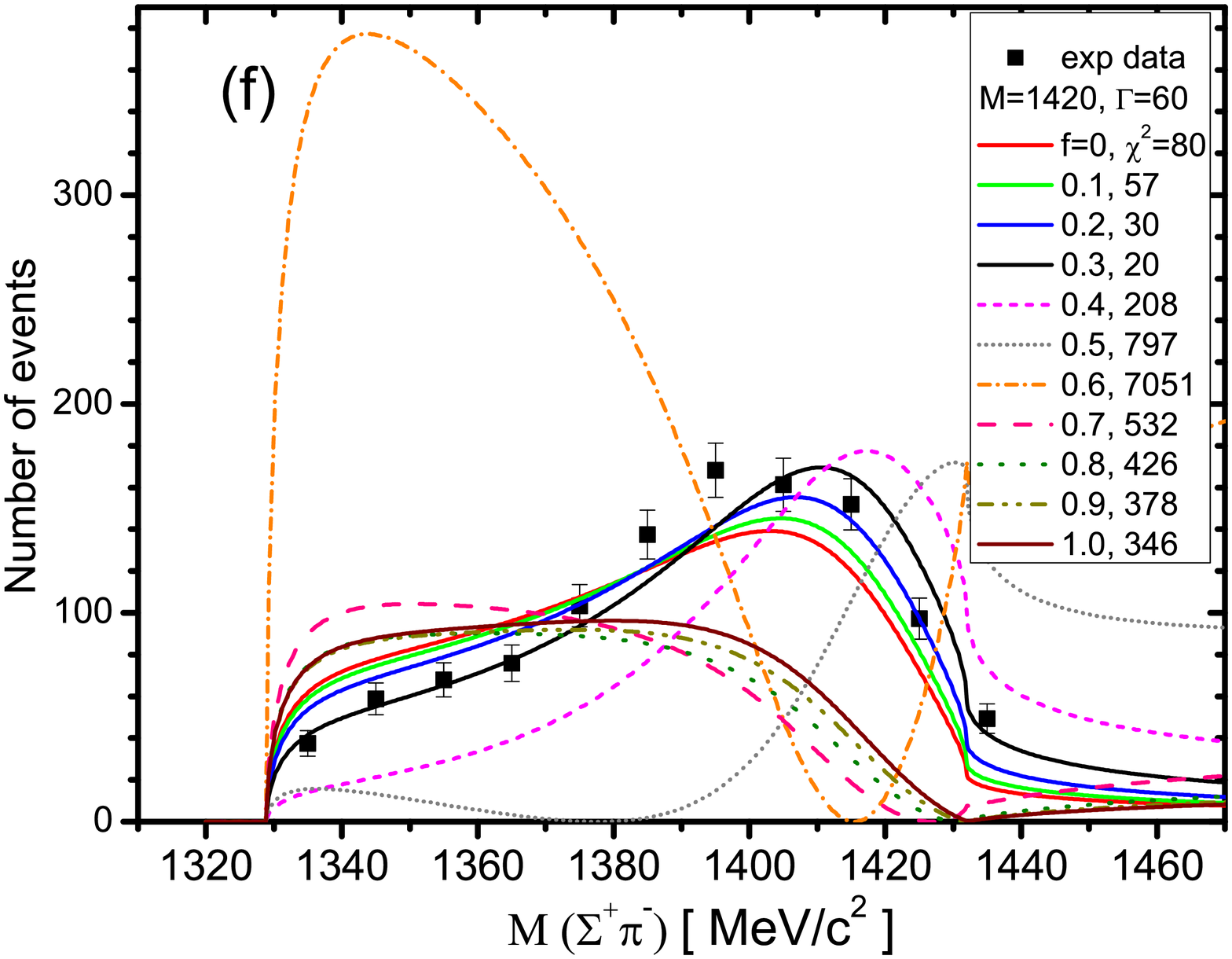}\\
\includegraphics[width=8cm]{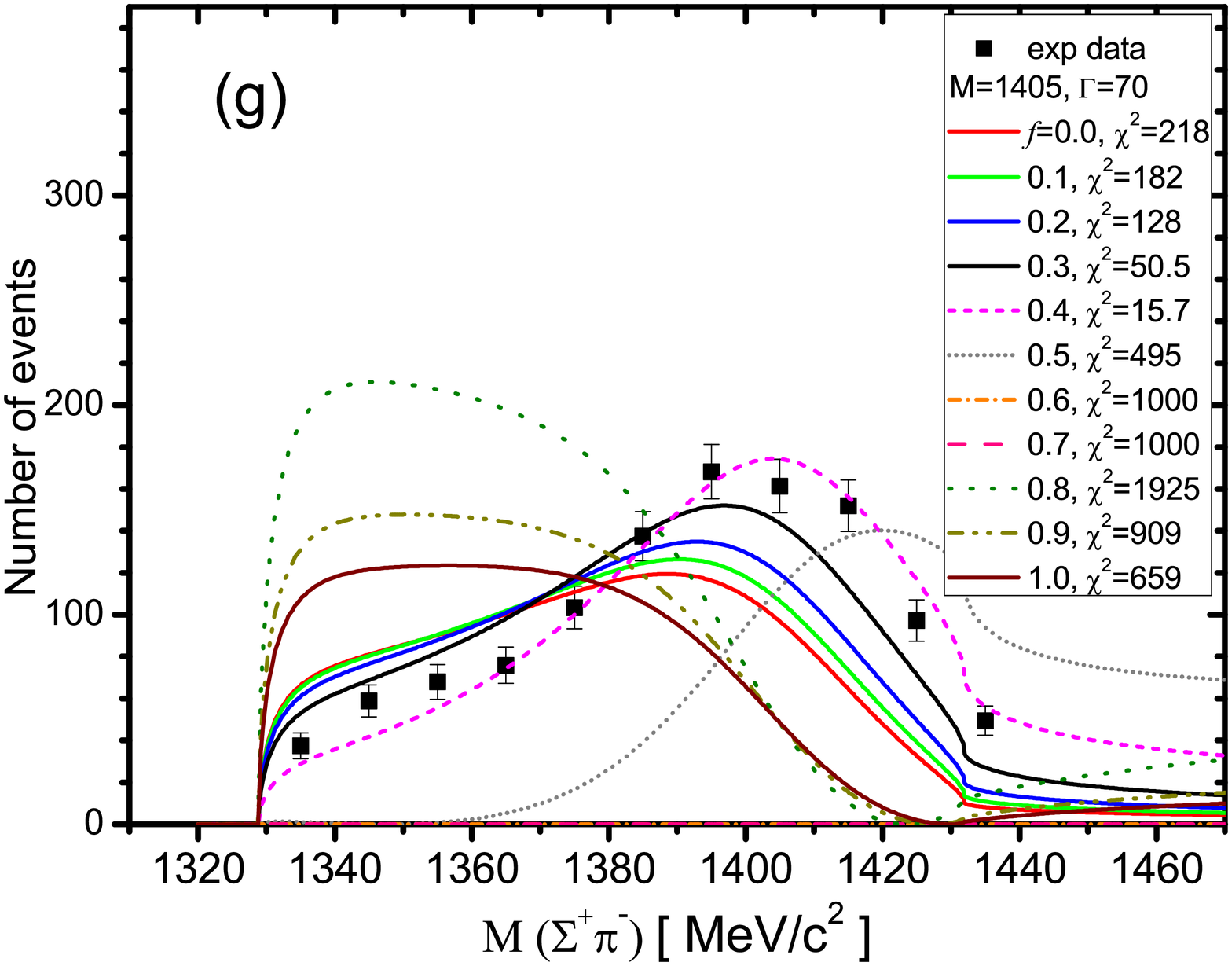}
\includegraphics[width=8cm]{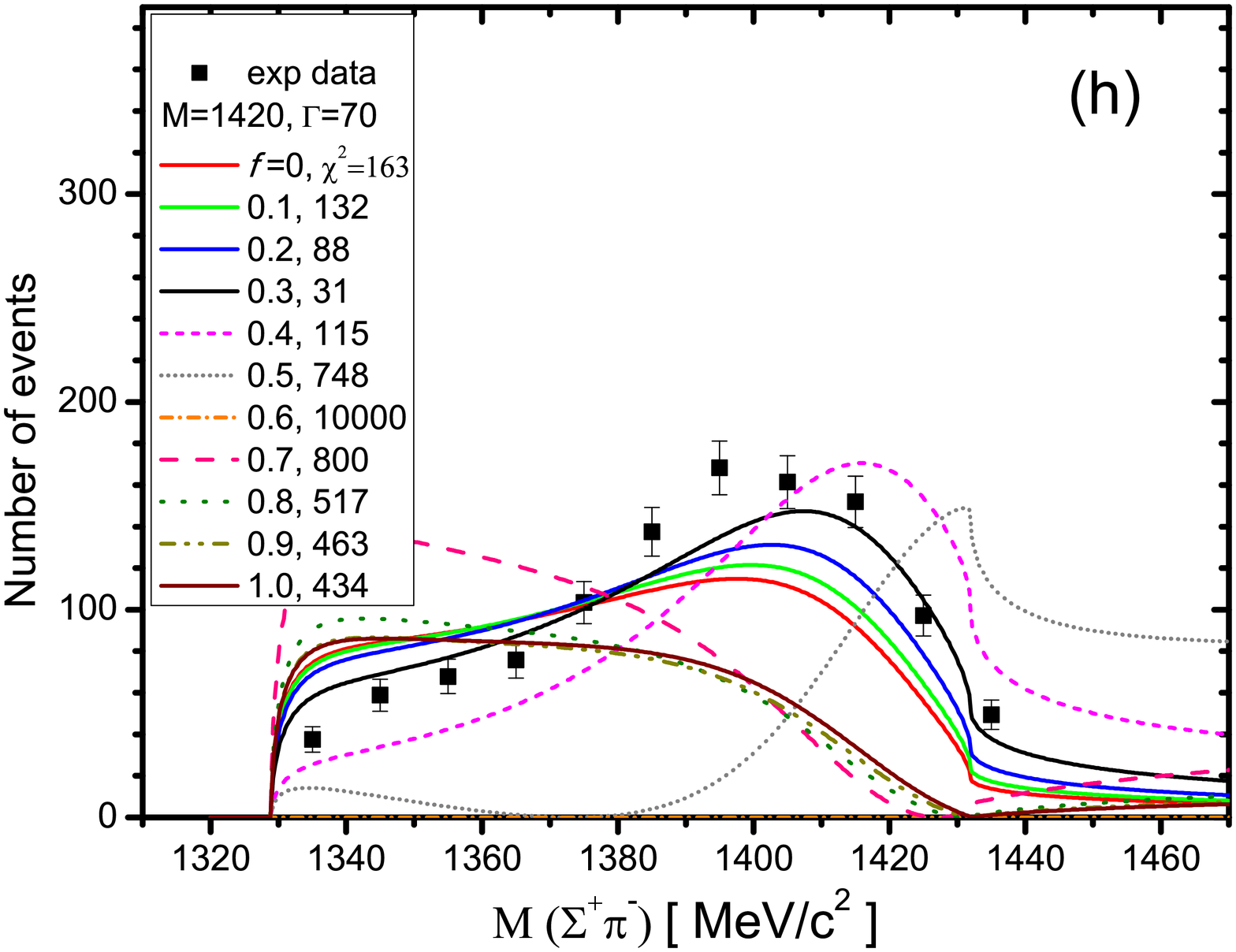}\\
\caption{(Color online) Calculated $M(\Sigma^+ \pi^-)$ spectra for mixed $f T_{21} + (1-f) T_{22}$ channels in which $\Gamma$ is fixed to various values and $M_{\Lambda ^*}$ is set to 1405 MeV/$c^2$ (left figures) and to 1420 MeV/$c^2$ (right figures). Each frame contains figures with mixing parameters $f = 0, 0.1,,,, 0.9,$ and 1.0 with best-fit curves and corresponding $\chi^2$ values. }\label{fig:T21+T22_All}
\end{figure*}

Figure \ref{fig:T21_Mfix_1405} shows typical $T_{21}$ and $T_{22}$ spectra with two assumed masses, $M$ = 1405 and 1420 MeV/$c^2$. The very broad character of the curves does not seem to allow good agreement with the experimental data at all. These figures show $\bar{K}N$ threshold effects on the $\Sigma \pi$ invariant mass spectrum, $|t_{21}|^2k_2$ and $|t_{22}|^2k_2$. When the width is sufficiently narrow, the spectrum is almost symmetric with a peak close to the assumed pole position \cite{Hassanvand13}. When the width becomes wide, the peak position is lowered from the pole position, and spectrum shape changes to a skewed one; this is the $\bar KN$ threshold effect on the spectrum. Although the $M$ value is assumed to be constant, the peak position shifts upon changing the $\Gamma$ value. In the case of a fixed $\Gamma$, as we change the $M$ value, the peak position shifts slightly from the pole position (compare the right frame of Fig. \ref{fig:T21_Mfix_1405} to the left one).

\begin{figure}
\center\includegraphics[width=9cm]{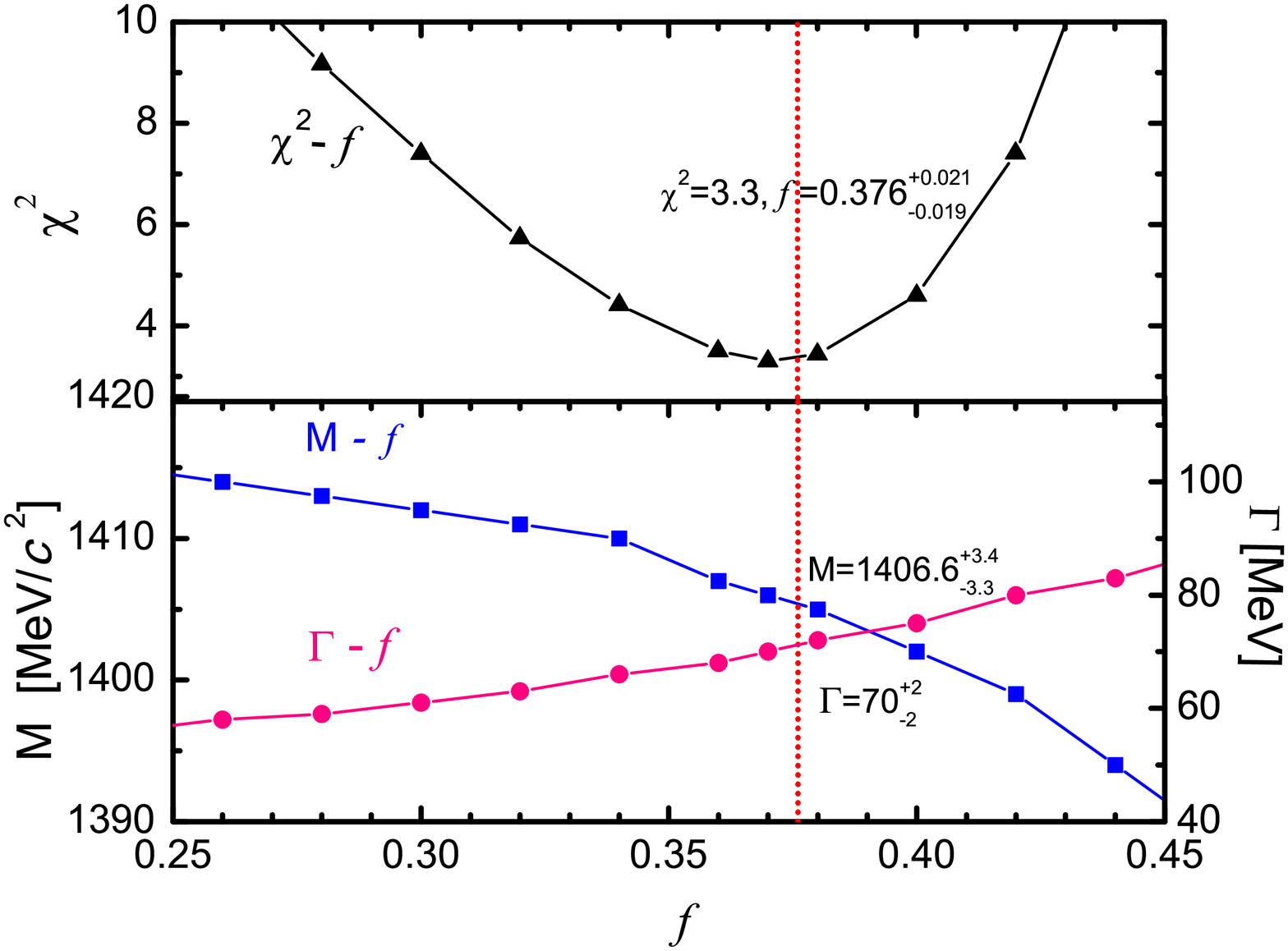}\\
\caption{(Color online) (Upper) $\chi^2$ versus $f$ and (Lower) $M$ and $\Gamma$ versus $f$. The best fit corresponds to $f = 0.376_{-0.019}^{+0.021}$, $M=1406.6_{-3.3}^{+3.4}$ MeV/$c^2$,  $\Gamma = 70 \pm 2$ MeV, with $\chi^2$=3.3 being shown by the vertical dotted line. }\label{fig:ML_f_chi2_G70}
\end{figure} 

\begin{figure}
\center\includegraphics[width=8cm]{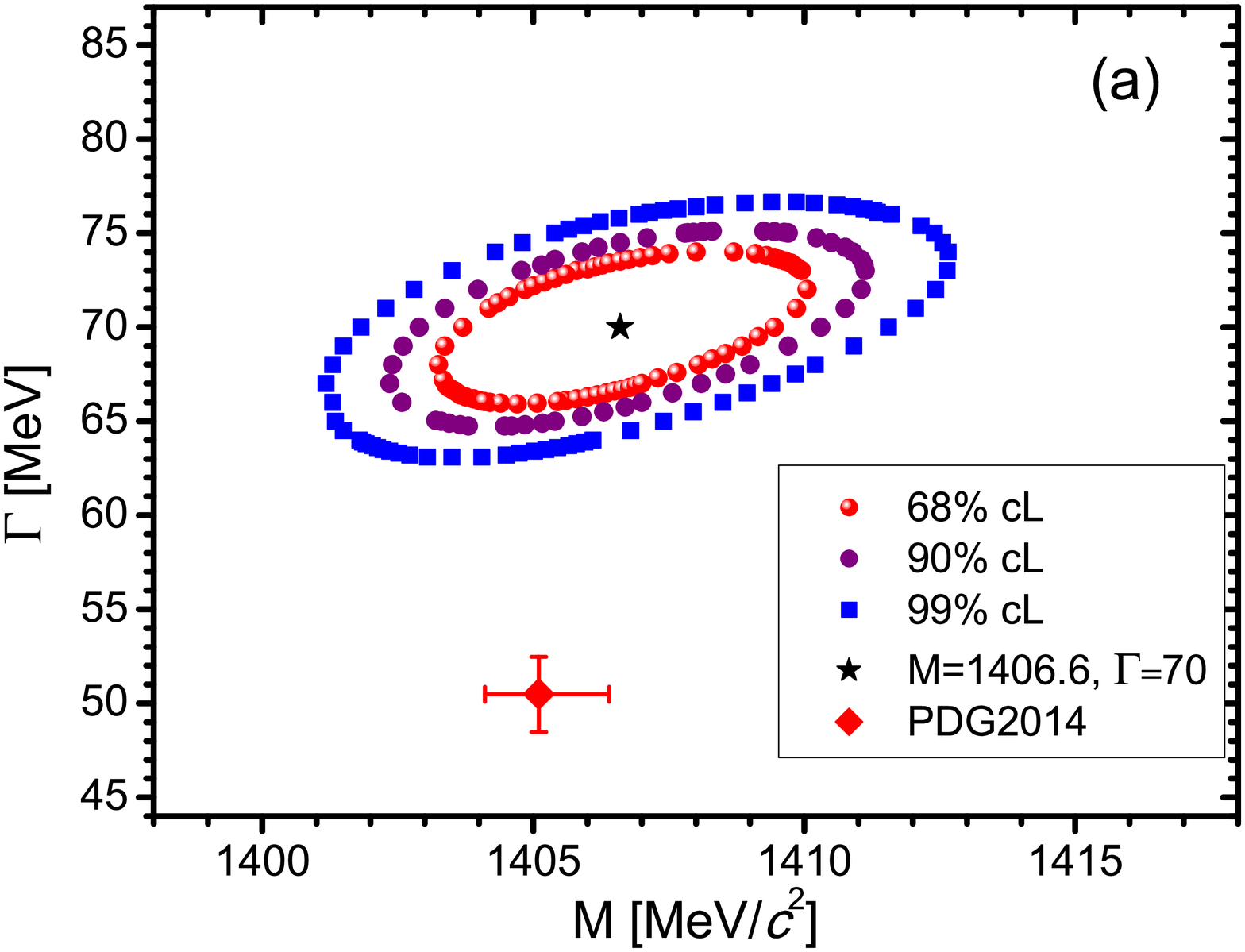}\\
\center\includegraphics[width=8cm]{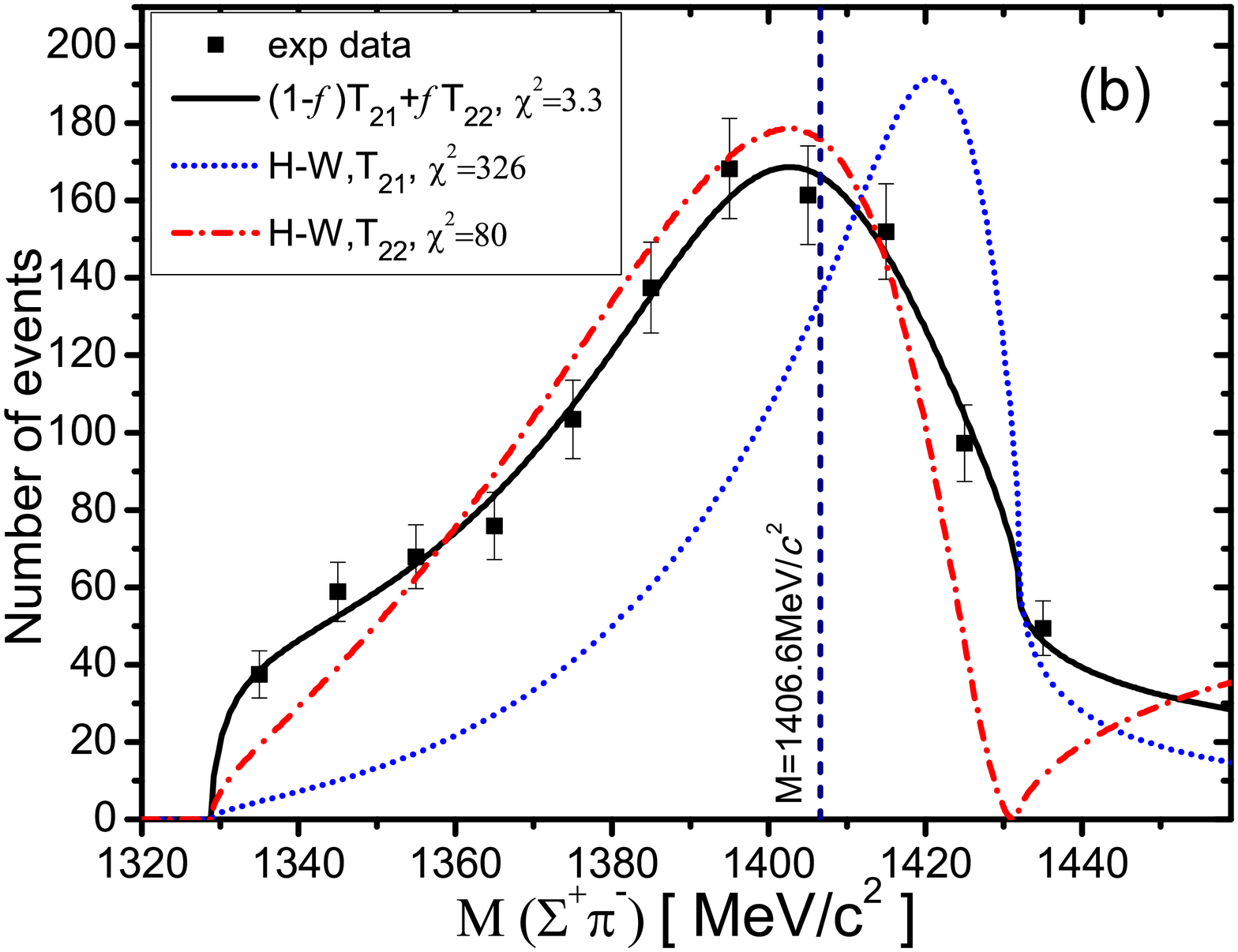}\\
\caption{(Color online) (Upper) Likelihood contour mapping of $M_{\Sigma \pi}$ vs $\Gamma$, where the best-fit values are obtained in a three-dimensional fitting of the Hemingway data with $(M, \Gamma, f)$. The results are: $M$ = $1406.6 _{-3.3}^{+3.4}$ MeV/$c^2$ , $\Gamma = 70 \pm 2 $MeV, $f = 0.376^{+0.021}_{-0.019}$, and $\chi^2 = 3.3$.  The contour curves are given for typical values of likelihood of 68, 95 and 99.9 percent.
(Lower) (Solid curve) The best-fit $M$ spectrum in our mixed $T_{21}$ and $T_{22}$ channel procedure. The present result is compared with individual $|T_{21}|^2k$ and $|T_{22}|^2k$ curves of Hyodo and Weise \cite{HW08}. The obtained mass from the present work is shown by the vertical dashed line.  }\label{fig:fig5}
\end{figure}

Because of the poor agreements between the experiment and theory using $T_{21}$ and $T_{22}$ alone, one might give up any fitting, but we now attempt to consider the case of mixed $T_{21}$ and $T_{22}$ transitions. This means that the shape of the spectrum is not produced from the $T_{21}$ or $T_{22}$ channel alone, but a mixture of both channels with different contributions is considered as follows. 
\begin{equation}
T_{\rm mixed} = (1-f) \,T_{21}+ f \,T_{22},
\end{equation}
with $f$ being a complex constant parameter. The percentages of $T_{21}$ and $T_{22}$ are $|1-f|^2/(|1-f|^2+|f|^2)$ and $|f|^2/(|1-f|^2+|f|^2)$, respectively.

 For this purpose, various combinations of the two channels were taken into account, and the shape of the spectra was plotted again. The situation is very complicated, since the vector addition of the complex functions $T_{21}$ and $T_{22}$ behaves in very strange ways. We show all the fitted curves with smoothly varying parameter values in Fig.~\ref{fig:T21+T22_All} which we show typical curves for $M =$1405 MeV/$c^2$ (left panels) and $M =$1420 MeV/$c^2$ (right panels) for different $\Gamma$ values of 40, 50, 60 and 70 MeV, in which the spectra were calculated for different values of the mixing parameter, $f$. Each frame contains 11 curves corresponding to 11 different mixing parameters of $f$, from 0 (pure $T_{21}$), 0.1, ,,,, to 1.0 (pure $T_{22}$). From a careful look at these figures we recognized that the mixing parameter between $f$ = 0.3 and 0.4 gives a strikingly better fitting of the experimental data. We thus made a three-dimensional fitting with three free parameters: $M$; $\Gamma$; $f$. 

Changing in $f$ alters the contribution of different channels, and the shape of the spectra is modified, where a shift of the peak position occurs. Once more, the best-fit process was iterated by varying $f$ smoothly.  
In fact, we were in a four-dimensional presentation of the $M-\Gamma-f-\chi^2$ parameters, where the minimum $\chi^2$ occurs only at one point; when $f$ comes close to a number of 0.376, which is equivalent to 27$\%$ of $T_{22}$ and 73$\%$ of $T_{21}$ contributions, the best result was obtained.

 Increasing the $f$ values causes a change in the shape of the spectrum, and makes the fit worse. In Fig.~\ref{fig:ML_f_chi2_G70} we plot $\chi^2$ versus $f$ (upper) and $M$ versus $f$ (lower). The parabolic behavior of the upper figure  shows a minimum value of 3.3 for $\chi^2$. 

Different complex numbers were checked for $f$, but real numbers produced a better outcome. 
 This significant shape of the spectrum is illustrated in Fig. \ref{fig:fig5} (lower), where Hemingway's data, our theoretical curve according to the best-fit parameters, and two curves of Hyodo-Weise's $T-$matrix calculation for $T_{21}$ and $T_{22}$ channels are combined together. This figure indicates that the spectra of the "chiral-weak" theory of Hyodo-Weise \cite{HW08} yield $\chi^2 = 326$ for the $T_{21}$ channel and $\chi^2 = 80$ for the $T_{22}$ channel, neither of which is in agreement with the experimental data. To make better sense of these results, the confidence level (CL) contours of $M$ versus $\Gamma$ are depicted at three levels of confidence  (68$\%$, 95$\%$ and 99.9$\%$) in Fig. \ref{fig:fig5} (upper). The most probable values, which correspond to the $1 \sigma$ uncertainty, are 

\begin{eqnarray}
f &=& 0.376^{+0.021}_{-0.019},\label{eq:fit-f}\\
M &=&  1406.6^{+3.4}_{-3.3} ~~{\rm MeV}/c^2,\label{eq:fit-M}\\
\Gamma  &=& 70 \pm 2 ~~{\rm MeV}, \label{eq:fit-G}
\end{eqnarray} 
with $\chi^2_{\rm MIN} = 3.3 $.

 The PDG 2014 values of  $M$ and $\Gamma$ with their error bars are also shown in the figure.

\section{Concluding remarks} 
The invariant-mass spectra of $\Lambda(1405) \rightarrow \Sigma \pi $ process in the decay of $\Sigma^+(1660)$ produced in the $K^-+p$ reaction at $4.2$ GeV were theoretically calculated and compared to experimental  data of Hemingway, which covers a range from the $ \Sigma+ \pi $ threshold (1330 MeV/$c^2$) to the $ K^-+p$ threshold (1430 MeV/$c^2$). Each spectrum shows a broad and skewed peak, reflecting both the $\Lambda^*$ pole and the lower and upper thresholds. The two different $T_{\Sigma\pi \leftarrow \bar{K} N}$ ($T_{21}$) and $T_{\Sigma\pi \leftarrow \Sigma\pi}$ ($T_{22}$) channels were taken into account, but neither of them showed good fits to the experimental spectrum. Then, a combination of the two channels was attempted, and significantly better fits were obtained with a $\chi^2$ minimum of 3.3. Finally, we obtained the best-fit values presented in (\ref{eq:fit-f},\ref{eq:fit-M},\ref{eq:fit-G}). 

This result shows that the $M$-value obtained from the Hemingway data is in good agreement with those from other old experimental data, summarized in Particle Data Group 2014, which further justifies the $\Lambda^*$ ansatz for deeply bound $\bar{K}$ nuclei, based on the strongly attractive $\bar{K}N$ interaction \cite{Akaishi:02,Yamazaki:02,Yamazaki:04,Dote:04a,Dote:04b}.

\section{Acknowledgments}

This work is supported by a Grant-in-Aid for Scientific Research of Monbu-Kagakusho of Japan. 
One of us (M.H.) wishes to thank A. Hassasfar for discussions during this work. 



\begin{thebibliography}{99}
\bibitem{Dalitz59} R.H. Dalitz and S.F. Tuan, Ann. Phys. {\bf 8}, 100 (1959).
\bibitem{L1405} M.H. Alston {\it et al.}, Phys. Rev. Lett. {\bf 6}, 698 (1961). 
\bibitem{PDG12} Particle Data Group 2012, J. Beringer {\it et al.},  Phys. Rev.  {\bf D 86}, 010001 (2012).
\bibitem{Dalitz-Deloff} R.H. Dalitz and A. Deloff, J. Phys. G: Nucl. Part. Phys. {\bf 17}, 289 (1991).
\bibitem{Hemingway} R.J. Hemingway, Nucl Phys. {\bf B 253}, 742 (1985). 
\bibitem{Akaishi:02}Y. Akaishi and T. Yamazaki, Phys. Rev.  {\bf C 65}, 044005 (2002).
\bibitem{Yamazaki:02}T. Yamazaki and Y. Akaishi, Phys. Lett. {\bf B 535}, 70 (2002).
\bibitem{Yamazaki:04} T. Yamazaki, A. Dot$\acute{\rm e}$, Y. Akaishi, Phys. Lett. {\bf B 587}, 167 (2004).
\bibitem{Dote:04a}A. Dot$\acute{\rm e}$, H. Horiuchi, Y. Akaishi and T. Yamazaki, Phys. Lett. {\bf B 590}, 51 (2004).
\bibitem{Dote:04b}A. Dot$\acute{\rm e}$, H. Horiuchi, Y. Akaishi and T. Yamazaki, Phys. Rev.  {\bf C 70}, 044313 (2004).
\bibitem{YA07PRC} T. Yamazaki and Y. Akaishi, Phys. Rev. {\bf 76}, 045201 (2007).
\bibitem{JidoNPA03} D. Jido, J.A. Oller, E. Oset, A. Ramos and U.G. Meissner, Nucl. Phys. {\bf A 725}, 181 (2003).
\bibitem{HW08}T. Hyodo and W. Weise, Phys. Rev.  {\bf C 77}, 035204 (2008).
\bibitem{Akaishi:10} Y. Akaishi, T. Yamazaki, M. Obu and M. Wada, Nucl., Phys. {\bf A 835}, 67 (2010). 
\bibitem{Kaplan-Nelson} D.B. Kaplan and A.E. Nelson, Phys. Lett. {\bf B 175}, 57 (1986).
\bibitem{Brown94} G.E. Brown, C.H. Lee, M. Rho and V. Thorsson, Nucl. Phys. {\bf A 567}, 937 (1994). 
\bibitem{Braun77} O. Braun {\it et al.}, Nucl. Phys. {\bf B 129}, 1 (1977).
\bibitem{Zychor08}I. Zychor {\it et al.}, Phys. Lett. {\bf B 660}, 167 (2008).
\bibitem{Esmaili10} J. Esmaili, Y. Akaishi and T. Yamazaki, Phys. Lett. {\bf B 686}, 23 (2010).
\bibitem{Esmaili11}J. Esmaili, Y. Akaishi and T. Yamazaki, Phys. Rev. {\bf C 83}, 055207 (2011).
\bibitem{Riley}B. Riley, I-T. Wang, J.G. Fetkovich and J.M. McKenzie, Phys. Rev. {\bf D 11}, 3065  (1975).
\bibitem{Hassanvand13}M. Hassanvand, S. Z. Kalantari, Y. Akaishi, T. Yamazaki, Phys. Rev. {\bf C 87}, 055202 (2013); Phys. Rev. {\bf C 88}, 019905(E) (2013).
\bibitem{HADES12b}G. Agakishiev {\it et al.} (HADES collaboration), Phys. Rev. {\bf C 87}, 025201 (2013).
\bibitem{PDG14} Particle Data Group 2014, K.A. Olive {\it et al.},  Chin. Phys. {\bf C 38}, 090001 (2014).
\bibitem{FINUDA} M. Agnello {\it et al.}, Phys. Rev. Lett. {\bf 94}, (2005) 212303.
\bibitem{DISTO} T. Yamazaki {\it et al.}, Phys. Rev. Lett. {\bf 104}, 132502 (2010); P. Kienle {\it et al.}, Eur. Phys. J. A {\bf 48}, 183 (2012).
\bibitem{E27} Y. Ichikawa {\it et al.}, Porg. Exp. Theor. Phys., 2015, 021D01 (2015). 
\bibitem{AMY08}Y. Akaishi, K. S. Myint and T. Yamazaki, Proc. Jpn. \\
Acad.  {\bf B 84}, 264 (2008).
\bibitem{Geng-Oset}L. S. Geng and E. Oset, Eur. Phys. J. {\bf A 34}, 405 (2007).
\bibitem{Feshbach58} H. Feshbach, Ann. Phys. {\bf 5}, 357 (1958); {\bf 19}, 287 (1962).
\bibitem{arXiv} M. Hassanvand {\it et al.}, Phys. Rev. C (2015), in press

\end{thebibliography}
\end{document}